\documentclass[aps,prl,reprint,superscriptaddress,longbibliography,showpacs]{revtex4-2}

\usepackage{graphicx}
\usepackage{amsmath}
\usepackage{amssymb}
\usepackage{bm}
\usepackage{array}
\usepackage{booktabs}   
\usepackage{multirow}    
\usepackage{longtable}   
\usepackage{dcolumn}     
\usepackage{amsmath}
\usepackage{xcolor}

\usepackage[normalem]{ulem}  
\usepackage[colorlinks,linkcolor=blue,citecolor=blue,urlcolor=blue]{hyperref}
\usepackage[capitalise]{cleveref}  

\usepackage{etoolbox}    
\usepackage{setspace}    
\usepackage{afterpage}   

\renewcommand{\sout}{\bgroup\color{red}\ULdepth=-.5ex\ULset}

\begin{document}

\title{{Nuclear Urca cooling in magnetar crust and anomalous heating}}

\author{Linjing Qi}
\affiliation{Key Laboratory of Beam Technology of Ministry
of Education, School of Physics and Astronomy, Beijing
Normal University, Beijing 100875, China}
\author{Qiye Hu}
\affiliation{School of Physical Science and Technology, Southwest University, Chongqing 400715, China}
\author{Xinxu Wang}
\affiliation{School of Physics, Peng Huanwu Collaborative Center for Research and Education, Beihang University, Beijing 100191, China}
\author{Yudong Luo}
\affiliation{Center for Exotic Nuclear Studies, Institute for Basic Science, Daejeon 34126, Korea}
\author{Ligang Cao}
\email{caolg@bnu.edu.cn}
\affiliation{Key Laboratory of Beam Technology of Ministry
of Education, School of Physics and Astronomy, Beijing
Normal University, Beijing 100875, China}
\affiliation{Institute of Radiation Technology,Beijing Academy of Science and Technology,Beijing 100875, China}
\author{Longjun Wang}
\email{longjun@swu.edu.cn}
\affiliation{School of Physical Science and Technology, Southwest University, Chongqing 400715, China}
\author{Toshitaka Kajino}
\email{kajino@buaa.edu.cn}
\affiliation{School of Physics, Peng Huanwu Collaborative Center for Research and Education, Beihang University, Beijing 100191, China}
\affiliation{Division of Science, National Astronomical Observatory of Japan, 2-21-1 Osawa, Mitaka, Tokyo 181-8588, Japan}
\affiliation{Graduate School of Science, The University of Tokyo, 7-3-1 Hongo, Bunkyo-ku, Tokyo 113-033, Japan}
\author{Hiroyuki Sagawa}
\affiliation{RIKEN Nishina Center, Wako 351-0198, Japan}
\affiliation{Center for Mathematics and Physics, University of Aizu, Aizu-Wakamatsu 965-8560, Japan}
\affiliation{Institute of Theoretical Physics, Chinese Academy of Sciences, Beijing 100190, China}
\author{Fengshou Zhang}
\affiliation{Key Laboratory of Beam Technology of Ministry of Education, School of Physics and Astronomy, Beijing Normal University, Beijing 100875, China}
\affiliation{Institute of Radiation Technology,Beijing Academy of Science and Technology,Beijing 100875, China}
\affiliation{Center of Theoretical Nuclear Physics, National Laboratory of Heavy Ion Accelerator of Lanzhou, Lanzhou 730000, China}
\begin{abstract}

We present a new result that the nuclear Urca process in magnetars, associated  with ultrastrong magnetic field, dramatically alters the neutrino cooling efficiency to support higher surface temperature, which is not common in ordinary neutron stars. A new formalism for neutrino luminosity accommodating finite temperature on top of magnetic field is extended with the use of realistic nuclear structure inputs for the Urca pair $^{29}$Na-$^{29}$Mg. 
We find that the neutrino luminosity of Urca pair shows a spiral variation as a function of magnetic field and crust density, \textcolor{black}{while the Urca pair persists}, in contrast to previous studies. Applying geometrically correct volumetric energy loss rate to Urca shell at fixed density with varying magnetic field strength, our results show that thermally populated excited states at finite temperature enhance neutrino luminosity, whereas conditions constrained Urca-pair formation driven by ultrastrong magnetic field reduce cooling efficiency, which can explain high temperature of magnetar surface as a new possibility.

\end{abstract}

\maketitle

\setlength{\parskip}{0pt}
 
  \textit{Introduction}---
  Strongly magnetized neutron stars, named magnetars, are among the most extreme stellar remnants in the Universe, with surface magnetic fields reaching well above 10$^{14}$ - 10$^{15}$ G \cite{Vigan2020,Turolla2015,Mereghetti2015}. 
  Their extreme magnetic field strength makes magnetars unique cosmic laboratories for probing physical processes far beyond the reach of terrestrial experiments \cite{Uzdensky2014,Farah2026,Potekhin2001,Kaspi2017}. 
\textcolor{black}{One of the unsolved questions in magnetar cooling is why magnetars exhibit systematically higher surface temperatures than ordinary neutron stars. X-ray spectroscopic observations show that magnetars maintain substantially higher thermal luminosities than ordinary neutron stars of comparable age }\cite{Potekhin2018,Grandis2020,Vigan2013,Potekhin2020-zm}, \textcolor{black}{with typical surface temperatures around 0.28 - 0.72 keV compared to 0.01 - 0.15 keV for their ordinary counterparts}  \cite{Maruyama2022,Perna2011}. \textcolor{black}{The prevailing interpretation attributes this anomalous heating to the dissipation of magnetic energy stored in the stellar interior \cite{Duncan1992,Pons2011,Li2016,Kaminker2006},
through mechanisms such as Joule heating from Ohmic decay \cite{Aguilera2008,Kaminker2006,Vigan2013}, ambipolar heating \cite{2023ApJ...945..151T,2023ApJ...945..151T,Beloborodov2016,Thompson2001-xb}, Hall drift \cite{Beloborodov2016,1992ApJ...395..250G}, magnetic field-decay electron captures \cite{Cooper2010-rs,Chamel2022} or additional heat sources in the outer crust \cite{Kaminker2006,Kaminker2007,Akgn2018,Ho2012}.} 
However, the physical origin of this persistent excess heating remains still unclear, constituting a long standing open problem in astrophysics 
\cite{Beloborodov2016,Turolla2015}. 

Urca process, a cyclic weak-interaction process of electron capture and $\beta^{-}$ decay,
emits neutrinos carrying away large amount of thermal energy 
\cite{Langanke2003,Suzuki2022,Yakovlev2004}. Although \textcolor{black}{``nucleon''} Urca process 
in neutron star core is shown to drive significant rapid cooling \cite{article,Lattimer1991,1994A&A...290..458H,Tsuruta1970,PhysRevLett.76.1994,Potekhin2026Urca}, the roles of \textcolor{black}{``nuclear''} Urca processes are generally ignored under zero-temperature approximation in crust models because Urca cycle is disfavored due to the blocking of electron phase space of $\beta^{-}$ decay \cite{Potekhin2015,Yakovlev2001}. 
Schatz $et$ $al.$  \cite{Schatz2013,2017PhDT24D} proposed that finite-temperature electron capture and $\beta^{-}$ decay transitions can establish an Urca cycle to operate within a thin shell of the accreting neutron-star crust, making the crustal nuclear Urca process an efficient source of neutrino cooling. Wang $et$ $al.$ \cite{Wang2021} subsequently demonstrated that the thermal population of low-lying excited states in both members of an Urca pair can further enhance the neutrino luminosity.


An Urca shell is also expected to form in the crust of magnetars.
As the magnetic field decays, the electron degeneracy pressure increases, associated with the increase of chemical potential $\mu_{e}$, which provides additional support for the crust against gravitational compression \cite{Cooper2010-rs}. Once chemical potential reaches the threshold energy $\lvert  Q_{EC} \rvert$ for exothermic electron captures, $\beta^{-}$ decay is activated at finite temperature, forming the Urca cycle.  Additionally, magnetar crustal magnetic field strength usually exceeds its surface field by one to two orders of magnitude in the regime above  critical value $B_{c}$ $\equiv$ ${m_{e}^{2}c^{3}}/{e\hbar}$ $\approx$ 4.414 $\times$ $10^{13}$ G
\cite{Geppert2004,Potekhin2018,wg2d-fx7j,Tambe2025,Lai1991ColdEO}. A recent analysis of the soft gamma-ray repeater
SGR 1806-20 inferred an internal toroidal magnetic-field component as strong as $B_{tor}$ $\sim$ 10$^{16}$ G and suggested that
fields of this magnitude may be common among magnetars 
\cite{Makishima2024-ll}.  At
such field strengths, Landau quantization of transverse electron motion modifies the electron density of states and, consequently, the phase space available for electron capture and $\beta^{-}$ decay
\cite{Konar2017,Potekhin2007,Cardall2001,Tambe2025,Famiano2022,Delsante1980,Lai1991ColdEO}. 
 The impact on crustal nuclear Urca cooling, nevertheless, remains largely unexplored. In this Letter, we present the first investigation of the combined effects of strong magnetic fields and finite temperatures on Urca cooling in magnetar crusts and their implication in higher thermal luminosities and surface temperature.

 \textit{Theoretical framework}---For the Urca shell, neutrino luminosity $L_{\nu}$ can be acquired through the integration of the total flux for neutrinos and anti-neutrinos $\mathcal{F}$ = $\mathcal{F}^{+}$ +
$\mathcal{F}^{-}$ \cite{Wang2021,Tsuruta1970} over the surface of a neutron star
\begin{equation}\label{Eq.1}
L_\nu = \int (\mathcal{F}^{+}+\mathcal{F}^{-})ds,
\end{equation}
where the superscript $+$ ($-$) denotes electron capture ($\beta^{-}$ decay). 
In the present work, instead of the previous approach of summing over all individual transitions at different position and density \cite{Wang2021,Sharma2025-vk}, we adopt the Urca total volumetric energy loss rate $ \mathcal{Q^{\pm}}$ by 
\begin{equation}\label{Eq.2}
 \mathcal{Q^{\pm}} =\tilde{L}{^\pm}(B)n^{\pm},
\end{equation}
where $\tilde{L}{^\pm}(B)$ is the total neutrino emissivity across all transitions under strong magnetic field \textcolor{black}{$B$} \cite{wg2d-fx7j} at the same position and density, including those of thermally populated excited states introduced at finite temperature \cite{Fuller1980-bu,Fuller1982-ex,Fuller1982-im,Fuller1985-sx}. $n^{\pm}$ denotes the total number density of parent nuclei. Then the neutrino and anti-neutrino fluxes $\mathcal{F}^{\pm}$ in Eq. (\ref{Eq.1}) reads
\begin{equation}\label{Eq.3}
 \mathcal{F}^{\pm}=\int_{R_{1}}^{R_{2}}\mathcal{Q^{\pm}}dr=\int_{R_{1}}^{R_{2}}\tilde{L}(B){^\pm}n^{\pm}dr,
\end{equation}
where $R_1$ and $R_2$ represent the boundary radii of the shell. $\mathcal{F}^{\pm}$ can be further rewritten employing the derivative of the Fermi energy with respect to the radius $r$ near the shell as
\begin{equation}\label{Eq.4}
 \mathcal{F}^{\pm}= \frac{1}{a_{0}}\int_{W_{F1}}^{W_{F2}}\tilde{L}(B){^\pm}n^{\pm}dW_{F}, 
\end{equation}
with $a_{0}$ \cite{Meisel2022-jk,Schatz2013,Wang2021} being
\begin{equation}\label{Eq.5}
 a_0 \equiv\frac{dW_{F}}{dr}= \frac{1}{10} \left( \frac{E_0}{m_e c^2} \right) \left( \frac{1}{\rho_9 Y_e} \right) \left( \frac{\rho g}{P_0} \right).
\end{equation}
Here $E_0$ = 3.7 MeV and $P_{0} = 3.6\times10^{26}$ erg cm$^{-3}$ stand for the energy per nucleon and the pressure of symmetric nuclear matter at saturation density  \cite{Meisel2022-jk,Schatz2013,Wang2021}. $\rho_9$ is the density $\rho$ in units of $10^{9}$ g cm$^{-3}$. $Y_e$ $\approx$ $Z/A$ is the electron fraction with $Z$ and $A$ being the charge and mass number of parent nucleus in electron capture for an Urca pair \cite{Wang2021,Deibel2016,Tsuruta1970}. $g$ = 1.85 $\times$ $10^{14}$ cm s$^{-2}$ is the surface gravity of neutron star  \cite{Meisel2022-jk,Schatz2013,Wang2021}. Our treatment is more physically realistic than the approach adopted in previous studies, which make the crude assumption \textcolor{black}{that the transitions among the ground and excited states of the parent and daughter nuclei occur at different densities and sites.} (See supplemental materials \cite{su} for more details). 

In the presence of magnetic field, the electron number density $n_e$ is related to Fermi energy $W_{F}$  through \cite{Delsante1980,7ytg-wzl8}
\begin{equation}\label{Eq.6}
n_e=\frac{2eB}{h^2c^2}\sum_{n=0}^{n_{max}}(2-\delta_{n0})[(W_{F}^2-(m_ec^2)^2-2neB\hbar c)^{1/2}],
 \end{equation}
   where $n$ and $n_{max}$ represent individual and maximum Landau level numbers, respectively. Standard notation are adopted for other constants \cite{Delsante1980,Lai1991ColdEO}. Based on the presumption of a single Urca pair within the Urca shell, under the joint conditions of steady-state equilibrium and detailed balance and charge neutrality \cite{Tsuruta1970,Wang2021}, $n^{\pm}$ can be rewritten as

{\footnotesize
 \begin{equation}\label{Eq.7}
n^{\pm}=\frac{2eB}{h^2c^2{Z}}\sum_{n=0}^{n_{max}}(2-\delta_{n0}) \sqrt{W_{F}^2-(m_ec^2)^2-2neB\hbar c}\frac{ \widetilde{\lambda}^\mp(B)}{\widetilde{\lambda}^+ (B)+ \widetilde{\lambda}^-(B)},
 \end{equation}
}where $\widetilde{\lambda}^\pm(B)$ indicates effective thermal decay rate with Boltzmann distribution weighting \cite{Fuller1980-bu,Fuller1982-ex,Fuller1982-im,Fuller1985-sx,wg2d-fx7j}. Using Eq. (\ref{Eq.4}) and Eq. (\ref{Eq.7}), $\mathcal{F}^{\pm}$ can be rewritten as

\begin{align}\label{Eq.8}
     \mathcal{F}^{\pm}
                           =&\frac{2eB}{h^2c^2a_{0}Z}\sum_{n=0}^{n_{max}}(2-\delta_{n0})  \int_{W_{F1}}^{W_{F2}}\\
                           &[(W_{F}^2-(m_ec^2)^2-2neB\hbar c)^{1/2}]
                        \frac{\widetilde{L}^{\pm} (B)\widetilde{\lambda}^{\mp}(B)}{\widetilde{\lambda}^{+}(B) + \widetilde{\lambda}^{-}(B)}dW_{F}.
\end{align}

\noindent Our expression features a geometrically continuous variation of realistic Urca cooling strength as a dependence of the Fermi energy, which
is the most important improvement compared with the previous analytic formulae \cite{Wang2021,Sharma2025-vk}.
\begin{figure}
\centering
	\includegraphics[width=\columnwidth,
    trim=0.07cm 0cm 0.15cm 0.0cm,clip]{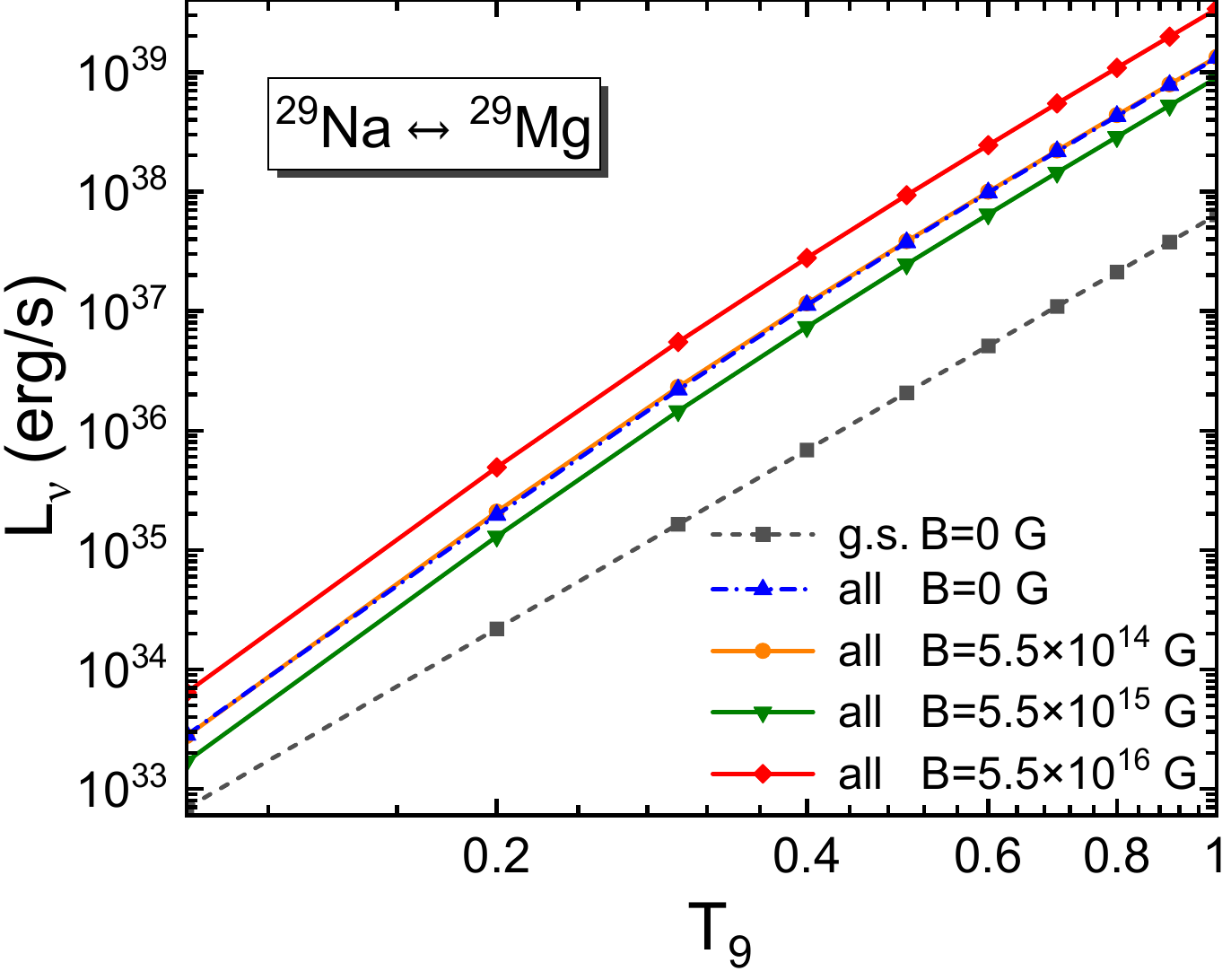}
	\caption{(color online) $L_{\nu}$ for Urca pair $^{29}$Na-$^{29}$Mg as a function of crust temperature.  Our predictions at 5.5$\times$10$^{14}$ G, 5.5$\times$10$^{15}$ G and 5.5$\times$10$^{16}$ G are shown as solid orange, green and red lines, respectively. For comparison, results in the absence of magnetic field effect for Ref.~\cite{Schatz2013} and  Ref.~\cite{Wang2021} are denoted as black short-dashed and blue dash-dotted lines. g.s. (all) labels ground-state transition (all transitions).}
	\label{fig 1}
\end{figure}

\textit{Discussion}---We use $^{29}$Na-$^{29}$Mg Urca pair as a demonstrative example to illustrate the mechanism of Urca process in  magnetar crust because this Urca pair is predicted to be the dominant source of neutrino luminosity in accreting neutron stars \cite{Schatz2013,2017PhDT24D}. The nuclear structure inputs used in our calculations are specified as follows. The thermally populated low-lying excited states at $E_{ex}$ $\lesssim$ 100 keV, i.e. the first excited state of $^{29}$Mg ($E_{ex}$ = 54 keV, $J^{\pi}=\frac{1}{2}^{+}$) \cite{https://doi.org/10.18139/nndc.ensdf/1845010,PhysRevC.36.765} and $^{29}$Na ($E_{ex}$ = 72 keV, $J^{\pi}=\frac{5}{2}^{+}$) \cite{https://doi.org/10.18139/nndc.ensdf/1845010,PhysRevC.36.765}, in addition to the ground states $^{29}$Mg ($J^{\pi}=\frac{3}{2}^{+}$) \cite{https://doi.org/10.18139/nndc.ensdf/1845010,PhysRevC.36.765} and $^{29}$Na ($J^{\pi}=\frac{3}{2}^{+}$) \cite{https://doi.org/10.18139/nndc.ensdf/1845010,PhysRevC.36.765}, and the transitions among them are taken into account. The log $ft$ value 5.06 is adopted for that transition between the ground states of  $^{29}$Na-$^{29}$Mg in both $\beta^{-}$ decay and electron capture \cite{PhysRevC.36.765}, while the log $ft$ values for the other channels of $\beta^{-}$ decay and electron capture are adopted from projected shell model calculations \cite{Wang2021,WOS:001568741400001}, $i.e.$ 6.24 (6.07) and 3.45 (3.15) for the ground state to excited state transition and its reverse, respectively. For the transition between the excited states, first-excited-state-to-first-excited-state, $\log ft=10.00$ ($\log ft=9.52$) is adopted as a reasonable estimate for a second forbidden transition in $\beta^{-}$ decay (electron capture) \cite{Wang2021}. 

We show in Fig. \ref{fig 1} the neutrino luminosity $L_{\nu}$ calculated by using our modified formula at the typical temperature range $T_{9}$ $\lesssim$ 1 (in units of $10^{9}$ K) with different strengths of zero and representative crust magnetic fields.  In our improved formulism of Eqs.~(\ref{Eq.1})-(\ref{Eq.5}), total volumetric energy loss rate $\mathcal{Q}$ is calculated over a narrow range of local densities  surrounding the original Urca location, where the all four transitions are evaluated consistently at the same physical location. Our approach accounts for the variation of the wide temperature range precisely by reducing the resultant $L_{\nu}$ by $\sim$ 13.4$\%$ systematically, in comparison with previous analytical treatment based on a uniform maximum-value approximation throughout the whole shell. 

  In Fig. \ref{fig 1}, for comparison, we show the calculated results without magnetic field by taking account of only the ground state (black short-dashed line) \cite{Schatz2013} and both ground and excited states (blue dash-dotted line) \cite{Wang2021}. $L_{\nu}$ in the latter case including excited states is larger than the former by more than one order of magnitude over the wide range of temperature, and its gradient becomes stronger. When magnetic field strength $B$ is turned on, $L_{\nu}$ values are more or less similar to one another among three cases with different field strength, $B$ = 5.5 $\times$ 10$^{14}$ G (orange solid line), 5.5 $\times$ 10$^{15}$ G (green solid line), and 5.5 $\times$ 10$^{16}$ G (red solid line). The orange solid line for $B$ = 5.5 $\times$ 10$^{14}$ G is almost the same as blue dash-dotted line without magnetic field. However, the $B$ dependence exhibits an interesting behavior: Although $L_{\nu}$ initially decreases with increasing $B$ (from the orange to the green solid lines), it rises again at stronger magnetic fields (from the green to the red solid lines). It suggests as if the neutrino luminosity $L_{\nu}$ oscillates as magnetic field $B$ changes. 
Our results in Fig. \ref{fig 1} reveal that strong magnetic fields fundamentally alter the Urca cooling, leading to a non-monotonic dependence of $L_{\nu}$ on magnetic-field strength, in sharp contrast to previous studies \cite{Schatz2013,Wang2021,Deibel2016,Sharma2025-vk,Meisel2022-jk}.

  \begin{figure*}
	\centering
	\includegraphics[width=11.8cm]{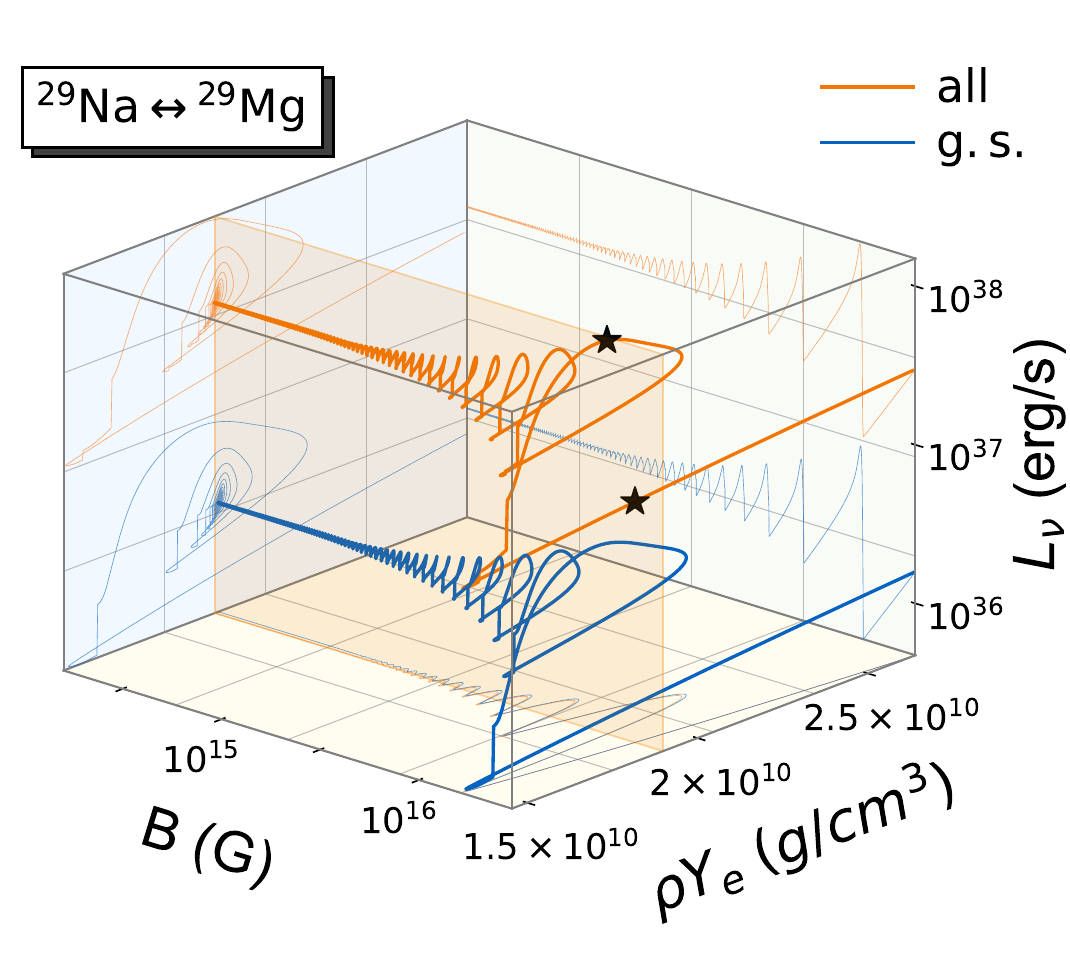}
	\caption{(color online) 
    $L_{\nu}$ for only ground state (blue line, g.s.) and all transitions (orange line, all) where Urca pair $^{29}$Na-$^{29}$Mg is formed at different magnetic field.
   $T_{9}$
= 0.5 is set to be the crust temperature. \textcolor{black}{Examples of $^{29}$Na-$^{29}$Mg  formed at $\rho Y_{e}$ $=$ 1.89 $\times$ $10^{10}$ g cm$^{-3}$ denoted as black stars.} }
	\label{fig 3}
  \end{figure*}
  
{
Fig. \ref{fig 3} illustrates the spiral trajectories which exactly satisfy the formation condition of the Urca cycle in magnetar-crust for a representative nuclear pair $^{29}$Na-$^{29}$Mg in three-dimensional space of the magnetic-field strength $B$, the electron density $\rho Y_{e}$ and the resultant neutrino luminosity $L_{\nu}$, i.e. $B$ - $\rho Y_{e}$ - $L_{\nu}$ space. 
The thick orange and blue curves show, respectively, the calculated results including all transitions among the ground and excited states of both \(^{29}\mathrm{Na}\) and \(^{29}\mathrm{Mg}\), and those considering only transitions between their ground states.
For each magnetic field strength $B$, the local density $\rho Y_{e}$ is chosen such that Urca pair satisfies the formation condition $\lvert Q_{EC} \rvert$ $-$ $k_{B}T$ $\lesssim$ $\mu_{e}$ $\lesssim$ $\lvert Q_{EC} \rvert$ $+$ $k_{B}T$. Here, $T_{9}$ = 0.5 is set to be the crust temperature and applied to all subsequent analyses. As shown in Fig. \ref{fig 3}, the thick orange curve lies approximately 1.2 orders of magnitude higher than the thick blue curve in $L_{\nu}$ for any values of $B$ and $\rho Y_{e}$, which indicates that thermally populated excited states dominate the neutrino luminosity $L_{\nu}$. At weak magnetic fields $B \lesssim$ 10$^{14.5}$ G, $L_{\nu}$ exhibits behavior nearly identical to that obtained with $B$ = 0 G.  However, once the magnetic field strength exceeds $B$ = 10$^{14.5}$ G, $L_{\nu}$ exhibits spiral-like oscillations with progressively increasing amplitude. Finally, for \(B \gtrsim 2 \times 10^{16}\,\mathrm{G}\), the spiral trajectory disappears, and \(L_{\nu}\) increases monotonically with increasing \(B\) and \(\rho Y_e\), while the Urca pair persists.
Note that their projections onto the $(B, L_{\nu})$, $(\rho Y_{e}, L_{\nu})$ and $ (B, \rho Y_{e})$ planes also are shown by the thin orange and blue curves in this figure for further discussions.}

Such a peculiar behavior of spiral trajectories at extremely strong magnetic field  can be understood as follows. Within the physical conditions considered here, magnetic field effects
are incorporated at its forming condition of Urca
pair, under the constraint of fixed chemical potential  satisfied $\lvert Q_{EC} \rvert$ $-$ $k_{B}T$ $\lesssim$ $\mu_{e}$ $\lesssim$ $\lvert Q_{EC} \rvert$ $+$ $k_{B}T$ in the crust. As shown on the projected yellow  $(B, \rho Y_{e})$ plane at the bottom of Fig.~\ref{fig 3}, since weak magnetic fields $B \lesssim$ 10$^{14.5}$ G give rise to high degeneracy of Landau levels and narrow energy separation $\hbar w_{c}$ ($w_{c}$=$\frac{eB}{m_{e}}$) between neighboring Landau tubes, electron density is almost constant in this regime. As the magnetic field strength increases, collisions among electrons become less effective at blurring quantized orbits, while the relatively low crustal temperature suppresses thermal excitation to adjacent Landau levels. This effect, therefore,  prevents the averaging-out of electron quantum oscillations.  These environmental conditions drive oscillation in electron density $\rho Y_{e}$ due to $\textbf{\textit{de-Haas-van-Alphen}}$ effect \cite{Kittel1954IntroductionTS}. When magnetic field strength further increases above \(B \gtrsim 2 \times 10^{16}\,\mathrm{G}\) to confine the system into the lowest Landau level, $\rho Y_{e}$ increases \textcolor{black}{quickly} with the increase of magnetic field. (See supplemental materials \cite{su} for more details). 

Both thin orange and blue lines in left blue $(\rho Y_{e}, L_{\nu})$ plane in Fig.~\ref{fig 3} reveal that, in contrast to the previous studies assuming a fixed density \cite{Schatz2013,Wang2021,Deibel2016,Sharma2025-vk,Meisel2022-jk}, the Urca pair can form for wider range of electron density $\rho Y_{e} \sim (1.5 - 2.6) \times 10^{10}$ g cm$^{-3}$ for any strength of the magnetic field $B$, even higher than the critical field strength \(B \sim 2 \times 10^{16}\,\mathrm{G}\) above which only the lowest Landau level is occupied.
Thus, the Urca pair including all transitions between the ground and excited states remains active over a much broader range of the magnetar crust under extremely strong magnetic field than that originally defined for accreting neutron star or white dwarf \cite{Schatz2013,Tsuruta1970}. 


We next explore the evolution of $L_{\nu}$ as a function of magnetic field $B$ in light-green projected $(B, L_{\nu})$ plane. 
Crust Urca cooling strength, which is proportional to volumertic energy loss rate, is characterized directly by the coherent behavior of neutrino emissivity {and electron density} (Eq. \ref{Eq.2}). As the magnetic field strength increases, the number of occupied Landau levels  below the Fermi surface $n_{max}$ decreases. Whenever the magnetic field reaches the critical value satisfying  ($n_{max}$ + $\frac{1}{2}$) $\hbar w_{c}$ = $W_{F}$, 
one Landau tube will be suddenly excluded from the Fermi sphere, resulting in a sudden drop in the neutrino emissivity and {electron density}, which corresponds to the minimum of $L_{\nu}$ of this plane. Between neighboring minimum points in $L_{\nu}$, $n_{max}$ remains unchanged, and the emissivity {as well as electron density} gradually increases with increasing magnetic field strength. Consequently, $L_{\nu}$ exhibits an overall zigzag structure dependence as a function of the magnetic field while the condition $\lvert Q_{EC} \rvert$ $-$ $k_{B}T$ $\lesssim$ $\mu_{e}$  $\lesssim$ $\lvert  Q_{EC} \rvert$ $+$ $k_{B}T$ is maintained. Therefore, a realistic picture of magnetar structure is required to discuss the cooling process due to variation of the neutrino luminosity.

Based on these analyses of the idealized cases in magnetar crust, we further investigate the behavior of $L_{\nu}$ as a function of magnetic field $B$ under the more realistic condition with fixed electron density. We emphasize here, again, that our $L_{\nu}$ in Eq. (\ref{Eq.1}), which takes account of volumetric energy loss rate Eq. (\ref{Eq.2}), is the sum of all transitions among the excited and ground states of $^{29}$Na and $^{29}$Mg at the same original site of Urca pair formation, although the previous studies \cite{Wang2021,Sharma2025-vk} virtually take rather unrealistic assumption that the individual transitions occur at different densities and sites. 

\begin{figure}
    \centering
	\includegraphics[width=\columnwidth,
    trim=0.07cm 0cm 0.15cm 0.11cm,clip]{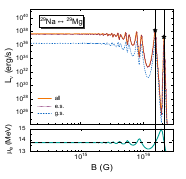}
	\caption{(color online) Upper and lower panels show $L_{\nu}$ and $\mu_{e}$ as a function of magnetic field at fixed $\rho Y_{e}$ $=$ 1.89 $\times$ $10^{10}$ g cm$^{-3}$, corresponding to the formation density of the Urca pair $^{29}$Na-$^{29}$Mg at $B$ = 0  G. Blue dashed, purple dash-dotted and orange solid lines represent the contributions from ground-state (g.s.), excited-state (e.s.), and all transitions, respectively. $T_{9}$
= 0.5 is set to be the crust temperature. \textcolor{black}{Black stars mark examples of $^{29}$Na-$^{29}$Mg  formed under extremely strong magnetic fields.}}
	\label{fig 2}
  \end{figure}

 Fig. \ref{fig 2} shows the evolution of neutrino luminosity $L_{\nu}$ as a function of magnetic field $B$ at the fixed value $\rho Y_{e}$ $=$ 1.89 $\times$ $10^{10}$ g cm$^{-3}$, which is the density where the Urca pair $^{29}$Na and $^{29}$Mg forms in the limit of $B$ $\Rightarrow$ 0. In this figure, the blue dashed-line, purple dotted-line and  orange solid-line display the contributions from only ground state transitions, the transitions including excited states, and their sum, respectively. Similarly to Fig. \ref{fig 3}, the excited states provide the dominant contribution to total $L_{\nu}$, while the contribution from the only ground states is suppressed by $\sim$ 1.2 orders of magnitude. $L_{\nu}$ also remains essentially unchanged from the zero-field value at lower magnetic fields $B \lesssim$ 10$^{14.5}$ G, 
and it oscillates with progressively larger amplitude as $B$ increases. 
Although both oscillations in Figs.~\ref{fig 3}-\ref{fig 2} are caused by de-Haas-van-Alphen effect,
the mechanisms are 
different from each other. (See supplemental materials \cite{su} for more details). The present  oscillation in $L_{\nu}$ of Fig. \ref{fig 2} is associated with the change of chemical potential $\mu_{e}$, where the system shifts into and out of the Urca window, $\lvert Q_{EC} \rvert$ $-$ $k_{B}T$ $\lesssim$ $\mu_{e}$  $\lesssim$ $\lvert  Q_{EC} \rvert$ $+$ $k_{B}T$, as $B$ increases, as clearly indicated by the oscillation of $\mu_{e}$ in the lower panel. When  $\mu_{e}$ takes the same value as that in the limit of $B$ $\Rightarrow$ 0, Urca condition revives and $L_{\nu}$ becomes local maximum. The positions of black stars, for example, satisfy this condition together with many other maxima for lower magnetic field, which correspond to the counterparts on the orange
plane of Fig. \ref{fig 3} at the same fixed density $\rho Y_{e}$ $=$ 1.89 $\times$ $10^{10}$ g cm$^{-3}$. The electron chemical potential $\mu_{e}$ oscillates in a period of $\Delta\frac{1}{B}$, {which stems from successive Landau-level crossings at the Fermi surface \cite{Kittel1954IntroductionTS}}. Finally, for very strong magnetic field \(B \gtrsim 2 \times 10^{16}\,\mathrm{G}\), $L_{\nu}$ monotonically decreases where only the lowest Landau level exists.
 
 During the evolution of magnetar, crust magnetic field strength will decay from the initial value as high as $\sim$ $10^{16}$ G \cite{Hoyos2008,Makishima2024-ll}.
 We found in Fig. \ref{fig 2} that $L_{\nu}$ shows generally lowering link trend, although there appear local maxima. Consequently, Urca cooling efficiency in magnetar crust declines in comparison with normal neutron stars, which prevents losing interior heat. Thus, the suppression of neutrino luminosity,  resulting from the quenching of the crustal Urca process, may provide a possible explanation for the observational fact that the magnetars exhibit higher thermal luminosities and effective surface temperature than those of the ordinary neutron stars \cite{Vigan2013,Kaminker2006}.

\textit{Summary}---
To summarize, the crust nuclear Urca process is important for understanding magnetar neutrino cooling, while previous theoretical studies generally eliminated Urca cycle in the crust and underestimated its contribution.  
In the present work we proposed, for the first time, that nuclear Urca cycle forms in magnetar crust in the mechanism due to the electron captures induced by magnetic-field decay and the $\beta^{-}$ decay  at finite temperature. Furthermore, we developed a new formulism such that neutrino emission associated with all transitions among the excited and ground states of Urca pair is estimated for the same electron density at the same site of the crust under ultrastrong magnetic field in a geometrically correct volumetric treatment. In our demonstrative numerical calculation of $^{29}$Na-$^{29}$Mg Urca pair, the nuclear structure inputs were provided by {experimental data combined with} projected shell model {calculations}. As a result, we found that Urca shell can exist for broad range of magnetic field and electron density, \textcolor{black}{exhibiting} a spiral pattern of neutrino luminosity, and that the low-lying excited states at finite temperature largely increase Urca neutrino luminosity. \textcolor{black}{Applying our formalism to realistic Urca-shell geometry at fixed density for strong magnetic field, }we found that Urca pair revives or disappears and that the neutrino luminosity averaged over magnetic field ultimately decreases cooling efficiency. This result can offer a novel interpretation of \textcolor{black}{still-unknown} mechanism for high surface temperature of magnetars. Our results provide a benchmark for magnetar cooling models and \textcolor{black}{open a new avenue} for studying magnetar thermal evolution and its internal structure. \textcolor{black}{Studies of magnetar cooling curves and the structural evolution of the crust and ocean, incorporating realistic chemical compositions and multiple Urca pairs, are currently underway.}

\textit{Acknowledgments}---  L. J. Qi thanks for the helpful discussion of A. Dohi, H. Huang, G. Colo and Q. L. Niu. 
L. G. Cao thanks for the helpful discussion of G. Colo and B. Haskell.
T. Kajino acknowledges the support in part by the National Key R \& D Program of China (2022YFA1602401) and the National Natural Science Foundation of China (No. 12335009 \& 12435010).  L. G. Cao and their group are partly supported by the National Natural Science Foundation of China under Grant Nos. 12675151, 12275025,  12635008, 12135004, the National Key R$\&$D Program of China under Grant No. 2023YFA1606401. H. Sagawa is supported by the JSPS Grant-in-Aid for Scientific Research (C) under Grant No. JP26K07079. Y. Luo is supported by the Institute for Basic Science369 of the Republic of Korea (Grant No. IBS-R031-D1).

\begin{acknowledgments}
 
\end{acknowledgments}


\bibliography{1st}

\end{document}


\title{Supplemental Material for ``Nuclear Urca cooling in magnetar crust and anomalous heating''}

\author{Linjing Qi}
\affiliation{Key Laboratory of Beam Technology of Ministry
of Education, School of Physics and Astronomy, Beijing
Normal University, Beijing 100875, China}
\author{Qiye Hu}
\affiliation{School of Physical Science and Technology, Southwest University, Chongqing 400715, China}
\author{Xinxu Wang}
\affiliation{School of Physics, Peng Huanwu Collaborative Center for Research and Education, Beihang University, Beijing 100191, China}
\author{Yudong Luo}
\affiliation{Center for Exotic Nuclear Studies, Institute for Basic Science, Daejeon 34126, Korea}
\author{Ligang Cao}
\email{caolg@bnu.edu.cn}
\affiliation{Key Laboratory of Beam Technology of Ministry
of Education, School of Physics and Astronomy, Beijing
Normal University, Beijing 100875, China}
\affiliation{Institute of Radiation Technology,Beijing Academy of Science and Technology,Beijing 100875, China}
\author{Longjun Wang}
\email{longjun@swu.edu.cn}
\affiliation{School of Physical Science and Technology, Southwest University, Chongqing 400715, China}
\author{Toshitaka Kajino}
\email{kajino@buaa.edu.cn}
\affiliation{School of Physics, Peng Huanwu Collaborative Center for Research and Education, Beihang University, Beijing 100191, China}
\affiliation{Division of Science, National Astronomical Observatory of Japan, 2-21-1 Osawa, Mitaka, Tokyo 181-8588, Japan}
\affiliation{Graduate School of Science, The University of Tokyo, 7-3-1 Hongo, Bunkyo-ku, Tokyo 113-033, Japan}
\author{Hiroyuki Sagawa}
\affiliation{RIKEN Nishina Center, Wako 351-0198, Japan}
\affiliation{Center for Mathematics and Physics, University of Aizu, Aizu-Wakamatsu 965-8560, Japan}
\affiliation{Institute of Theoretical Physics, Chinese Academy of Sciences, Beijing 100190, China}
\author{Fengshou Zhang}
\affiliation{Key Laboratory of Beam Technology of Ministry of Education, School of Physics and Astronomy, Beijing Normal University, Beijing 100875, China}
\affiliation{Institute of Radiation Technology,Beijing Academy of Science and Technology,Beijing 100875, China}
\affiliation{Center of Theoretical Nuclear Physics, National Laboratory of Heavy Ion Accelerator of Lanzhou, Lanzhou 730000, China}

\date{\today}


\maketitle




This supplemental material describes our research method, computational framework, and numerical verification of results through two sections: 


\section{de-haas-van-alphen oscillations in electron density and chemical potential}

De-haas-van-alphen oscillations arise from Landau quantization of electron orbital energy levels in magnetic field with the condition of strong magnetic field and low temperature \cite{Kittel1954IntroductionTS}. Identical increments $\Delta\frac{1}{B}$ lead to the recurrence of degenerate orbitals \cite{Kittel1954IntroductionTS}. When magnetic field strength increases, the electron states degeneracy $D$ for each Landau level become larger:
\begin{figure}[htbp]
   
    \begin{minipage}{0.49\textwidth}
        \centering
\includegraphics[width=\linewidth,trim=0.6cm 0cm 2.cm 1.2cm,clip]{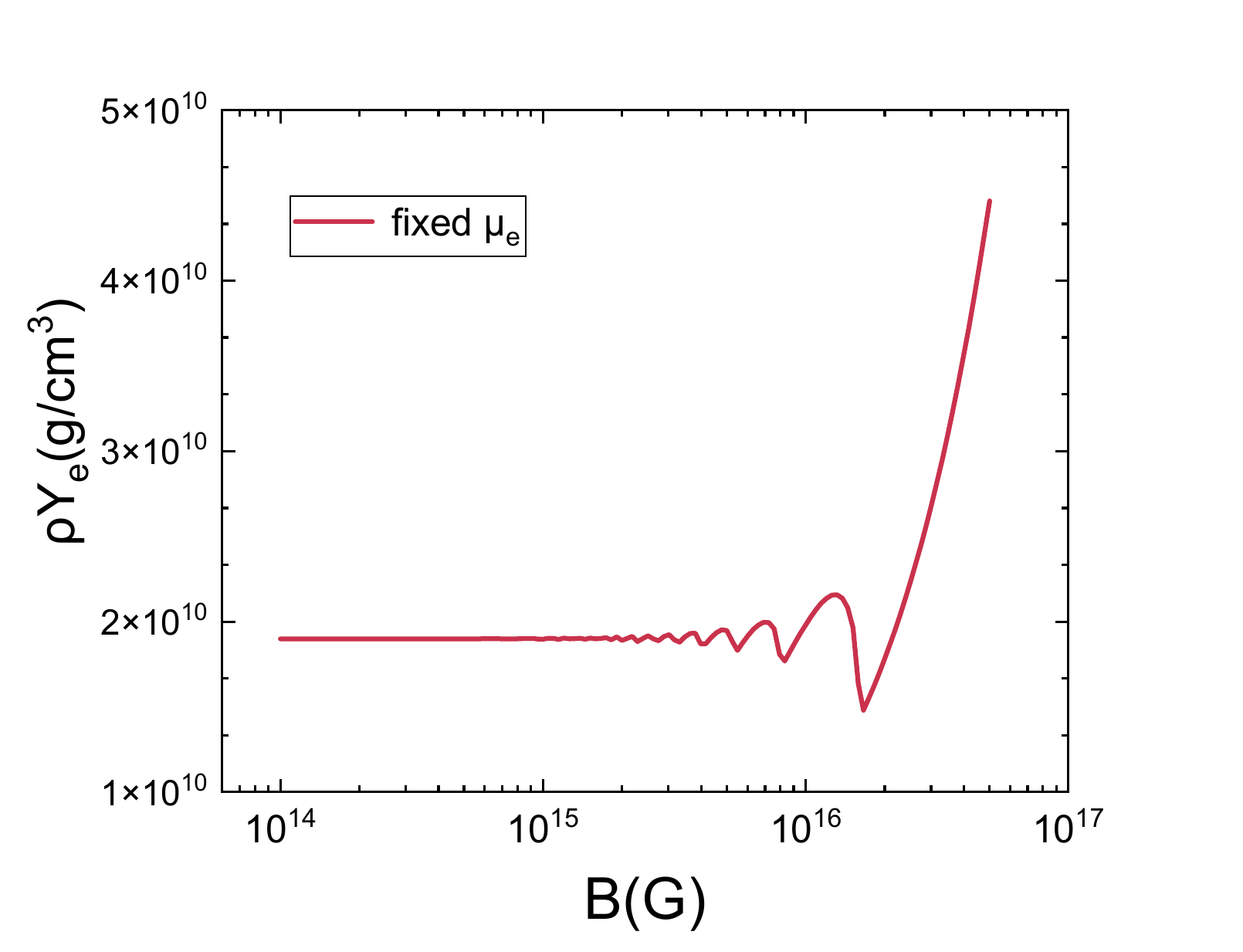}
        \caption{(color online) $\rho Y_{e}$ as a function of magnetic field at fixed $\mu_{e}$ where Urca pair $^{29}$Na-$^{29}$Mg forms with temperature $T_{9}$
= 0.5.}
        \label{fig 1}
    \end{minipage}
    \hfill
    \begin{minipage}{0.49\textwidth}
        \includegraphics[width=\linewidth,trim=0.6cm 0cm 2.cm 0.0cm,clip]{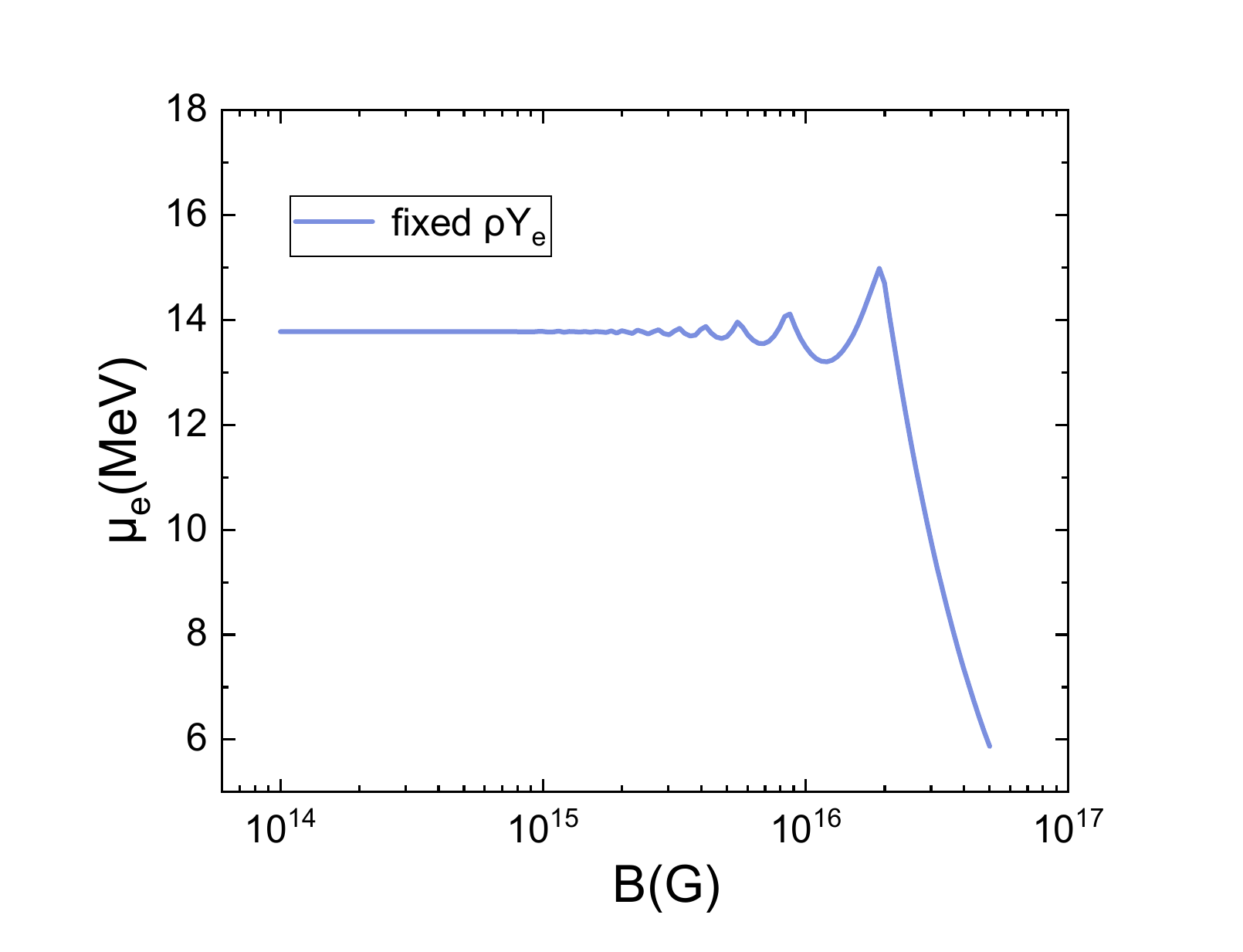}
 \caption{(color online) $\mu_{e}$ as a function of magnetic field at fixed $\rho Y_{e}$ = 1.89 $\times$ $10^{10}$ g cm$^{-3}$, corresponding to the  Urca pair $^{29}$Na-$^{29}$Mg formation density at $B$ = 0 G with temperature $T_{9}$
= 0.5.}
    \label{fig 2}
    \end{minipage}
\end{figure}

Case I (With fixed chemical potential): As is shown in Fig. \ref{fig 1}, Landau level successively vanishes when ($n_{max}$+$\frac{1}{2}$) $\hbar w_{c}$ = $W_{F}$. In the region between two critical magnetic-field value, electron density will increase with the increase of magnetic field.

 Case II (With fixed electron density): As is shown in  Fig. \ref{fig 2}, electron will suddenly jump to the neighboring lower Landau level when ($n_{max}$+$\frac{1}{2}$) $\hbar w_{c}$ = $W_{F}$ with the existing Landau levels within Fermi sphere satisfying ($n$-1) $D$ = $N$ ($N$ is the number of electrons). In the region between adjacent critical magnetic field, Fermi energy will decrease with the increase of magnetic field.

\section{Neutrino luminosity formula derivation}
\subsection{weak interaction rates with thermal nuclear excitation under strong magnetic field}
The effective thermal decay rates under strong magnetic field $\widetilde\lambda^{\pm}(B)$ and the corresponding neutrino energy loss rates $\widetilde L^{\pm}(B)$ can be expressed as \cite{Paxton2010,Fuller1980-bu,Fuller1982-ex,Fuller1982-im,Fuller1985-sx}
\begin{equation}\label{Eq.1}
\widetilde\lambda^{\pm}(B)=\frac{1}{G(Z,A,T)}\sum_{if}(2J_{i}+1)e^{-\frac{E_{i}}{kT}}\lambda^{\pm}_{if}(B),
 \end{equation} 
 \begin{equation}\label{Eq.2}
\widetilde L^{\pm}(B)=\frac{1}{G(Z,A,T)}\sum_{if}(2J_{i}+1)e^{-\frac{E_{i}}{kT}}L^{\pm}_{if}(B).
 \end{equation} 
Here $\pm$ denotes electron capture (+) or $\beta^{-}$ decay (-), respectively. $G(Z,A,T)$ = $\sum_{i}$(2$J_{i}$+1)exp(-$E_{i}$/$kT$) is the partition function of the parent nucleus with $J_{i}$ ($J_{f}$) and  $E_{i}$ ($E_{f}$) being the angular momenta and excitation energies for the initial state $i$ (final state $f$). $k_{B}T$ is the environmental temperature. The individual decay rates $\lambda^{\alpha}_{if}$ and corresponding neutrino energy loss rates $L^{\alpha}_{if}$ is given by
\begin{equation}\label{Eq.3}
\lambda^{-}_{if}(B)=exp({-\pi}\alpha Z)\frac{ln2}{ft^-_{if}}\frac{eB\hbar c}{2}\frac{1}{({m_ec^{2})}^{5}}\sum_{n=0}^{N_{max}}(2-\delta_{n0})\int_{\omega_{\beta}}^{Q_{if}}\frac{E(Q_{if}-E)^2}{\sqrt{E^2-m_e^2c^4-2neB\hbar c}}(1-S_{e})dE,
\end{equation}
 \begin{equation}\label{Eq.4}
\lambda^{+}_{if}(B)=exp({-\pi}\alpha Z)\frac{ln2}{ft^+_{if}}\frac{eB\hbar c}{2}\frac{1}{({m_ec^{2})}^{5}}\sum_{n=0}^{N_{max}}(2-\delta_{n0})\int_{\omega_{EC}}^{\infty}\frac{E(Q_{if}+E)^2}{\sqrt{E^2-m_e^2c^4-2neB\hbar c}}S_{e}dE,
\end{equation}
\begin{equation}\label{Eq.5}
 L^{-}_{if}(B)=m_{e}c^2exp({-\pi}\alpha Z)\frac{ln2}{ft^-_{if}}\frac{eB\hbar c}{2}\frac{1}{({m_ec^{2})}^{6}}\sum_{n=0}^{N_{max}}(2-\delta_{n0})\int_{\omega_{\beta}}^{Q_{if}}\frac{E(Q_{if}-E)^3}{\sqrt{E^2-m_e^2c^4-2neB\hbar c}}\times(1-S_{e}) dE,
\end{equation}
 \begin{equation}\label{Eq.6}
L^{+}_{if}(B)=m_{e}c^2exp({-\pi}\alpha Z)\frac{ln2}{ft^+_{if}}\frac{eB\hbar c}{2}\frac{1}{({m_ec^{2})}^{6}}\sum_{n=0}^{N_{max}}(2-\delta_{n0})\int_{\omega_{EC}}^{\infty}\frac{E(Q_{if}+E)^3}{\sqrt{E^2-m_e^2c^4-2neB\hbar c}}\times S_{e} dE,
\end{equation}
where  
 \begin{equation}\label{Eq.7}
\omega_{\rm EC} \equiv \max\left[-Q_{if}{},\, m_ec^2\right],
\end{equation}
\begin{equation}\label{Eq.8}
\omega_\beta \equiv \sqrt{m_e^2c^4 + 2neB\hbar c}\ ,
\end{equation}
\begin{equation}\label{Eq.9}
N_{\rm max} = \frac{Q_{if}^2 - m_e^2c^4}{2eB\hbar c}\ .
\end{equation}
Within Eq. (\ref{Eq.3},\ref{Eq.4},\ref{Eq.5},\ref{Eq.6}), 
$E$ = $\sqrt{p_{z}^{2}c^2+m_{e}^2c^4+2neB\hbar c}$ is the electron energy, where $p_{z}$ denotes the electron momentum component parallel to the magnetic field. $Q_{if}$ =  $M_{p}c^2$ $-$ $M_{d}$$c^{2}$ $+$ $E_{i}$ $-$ $E_{f}$ is the energy difference with $M_{p}(M_{d})$ representing nuclear mass of the parent (daughter) nucleus. The chemical potential $\mu_{e}$ in Eq. (\ref{Eq.3},\ref{Eq.4},\ref{Eq.5},\ref{Eq.6}) can be obtained by \cite{Delsante1980}
 \begin{equation}\label{Eq.10}
\rho Y_{\mathrm{e}}=\frac{2eB}{h^2cN_A}\sum_{n=0}^{\infty}\left(2-\delta_{n 0}\right)\int_0^{\infty}\left(S_{\mathrm{e}}-S_{\mathrm{p}}\right)dp,
 \end{equation}
where $\rho Y_{\mathrm{e}}$ is the electron density. $N_A$ is the Avogadro constant. Electron and positron Fermi-Dirac functions $(S_{e/p})$ can be expressed as \cite{Delsante1980}
\begin{equation}\label{Eq.11}
S_{e/p}(E) = \frac{1}{1+\exp(\frac{E \mp \mu_e}{kT})}.
\end{equation}

In the present work, the comparative half-life $ft_{if}^{\alpha}$ for the low-lying excited states are obtained by projeted shell model with $\frac{ft_{if}^{EC}}{2J^{EC}+1}$= $\frac{ft_{if}^{\beta ^{-}}}{2J^{\beta ^{-}}+1}$ \cite{Paxton2010} adopted a quenching factor of 0.75 for the reduced probability of nuclear transitions, except for the $ft$ for the ground state is adopted from experiment in Ref. \cite{PhysRevC.36.765}. The excitation energies and spin-parity assignments are taken from evaluated data where available (ENSDF) \cite{https://doi.org/10.18139/nndc.ensdf/1845010}.
\subsection{Urca neutrino luminosity}
The electron number density and Ferimi energy in magnetic field are related through \cite{Delsante1980,7ytg-wzl8}
\begin{equation}\label{Eq.12}
n_e=\frac{2eB}{h^2c^2}\sum_{n=0}^{n_{max}}(2-\delta_{n0})[(W_{F}^2-(m_ec^2)^2-2neB\hbar c)^{1/2}].
 \end{equation} 
Under the assumption of that there is only one pair in the Urca shell, further considering the condition of steady-state, detailed balance and charge neutrality \cite{Tsuruta1970,Wang2021} 
\begin{equation}\label{Eq.13}
\widetilde\lambda^{+}(B)n^+ = \widetilde\lambda^{-}(B)n^-,
\end{equation}
\begin{equation}\label{Eq.14}
n_e = Z n^+ + (Z - 1) n^-,
\end{equation}
the total number density for the parent nucleus can be rewritten as
 \begin{equation}\label{Eq.15}
n^{\pm}=\frac{2eB}{h^2c^2{Z}}\sum_{n=0}^{n_{max}}(2-\delta_{n0})(W_{F}^2-(m_ec^2)^2-2neB\hbar c)^{1/2}\frac{ \widetilde{\lambda}^\mp(B)}{\widetilde{\lambda}^+(B) + \widetilde{\lambda}^-(B)}.
 \end{equation}
Then the fluxes $\mathcal{F}^{\pm}$ can be expressed as 
\begin{align}\label{Eq.16}
     \mathcal{F}^{\pm}&=\int_{R_{1}}^{R_{2}}\tilde{L}{^\pm}(B)n^{\pm}dr
                           =\frac{1}{a_{0}}\int_{W_{F1}}^{W_{F2}}\tilde{L}(B){^\pm}n^{\pm}dW_{F}\\
                           &=\frac{2eB}{h^2c^2a_{0}Z}\{\sum_{n=0}^{n_{max}}(2-\delta_{n0})\int_{W_{F1}}^{W_{F2}}[(W_{F}^2-(m_ec^2)^2-2neB\hbar c)^{1/2}]\frac{\widetilde{L}^{\pm}(B) \widetilde{\lambda}^{\mp}(B)} {\widetilde{\lambda}^{+}(B)  + \widetilde{\lambda}^{-}(B) }dW_{F}\},
 \end{align}

\noindent with $a_{0}=dW_{F}/dr$ ($W_{F}$ $\approx$ $\mu_{e}$) being the derivative of the Fermi energy to the radius near the shell.
Here \cite{Wang2021,Tsuruta1970} \begin{equation}
 a_0 = \frac{1}{10} \left( \frac{E_0}{m_e c^2} \right) \left( \frac{1}{\rho_9 Y_e} \right) \left( \frac{\rho g}{P_0} \right), 
\end{equation}

\noindent where $E_0$ = 3.7 MeV and $P_{0} = 3.6\times10^{26}$ erg$ $ cm$^{-3}$ denote the energy per nucleon and pressure of symmetric nuclear matter at saturation density, respectively. $\rho_9$ is the density $\rho$ in units of $10^{9}$ g cm$^{-3}$. $Y_e$ $\approx$ $Z/A$ is the electron fraction with $A$ being the mass number of Urca pair. $g$ = 1.85$\times$ $10^{14}$ cm s$^{-2}$ is the surface gravity of neutron star.

The neutrino luminosity can be obtained by integrating the flux over the surface of a neutron star with radius $R$ \cite{Deibel2016,Schatz2013},
\begin{align}\label{Eq.17}
L_\nu &= \int \mathcal{F} ds \approx 4\pi R^{2} \mathcal{F}=4\pi R^{2}    
 (\mathcal{F}^{+}+\mathcal{F}^{-}).
\end{align}

\noindent In the present work, $R$ = 12 km is adopted as the typical neutron star radius \cite{Deibel2016,Schatz2013}

.

\bibliography{supplement2/supplementsNotes} 